\documentclass[prd,twocolumn,showpacs,floatfix,amsmath,nofootinbib,amssymb,floatfix]{revtex4}
\usepackage{graphicx,color,dcolumn,booktabs,bm}
\usepackage{longtable,lscape}
\usepackage{txfonts}
\usepackage{overpic}
\usepackage{amssymb}
\usepackage{amsmath}
\usepackage{indentfirst}
\usepackage{feynmf}   
\usepackage{slashed}  
\usepackage{cases}
\usepackage{color}
\usepackage{multirow}
\usepackage{epstopdf}
\usepackage{ulem}
\usepackage{subfigure}
\usepackage[separate-uncertainty]{siunitx}

\newcommand{\eref}[1]{Eq.~(\ref{#1})}
\newcommand{\w}{\omega}
\newcommand{\vo}{\vec{o}\@ifnextchar{^}{\,}{}}
\def\up{\mathrm}

\newcommand{\vabs}[1]{|\vec #1\,|}

\graphicspath{{PDF/}}
\def\dx{\!\cdot\!}
\def\slash#1{\setbox0=\hbox{$#1$}           
	\dimen0=\wd0                                 
	\setbox1=\hbox{/} \dimen1=\wd1               
	\ifdim\dimen0>\dimen1                        
	\rlap{\hbox to \dimen0{\hfil/\hfil}}      
	#1                                        
	\else                                        
	\rlap{\hbox to \dimen1{\hfil$#1$\hfil}}   
	/                                         
	\fi}                                         %
\def\sl#1{\setbox0=\hbox{#1}
	\dimen0=\wd0
	\rlap{\hbox to \dimen0{\hss/\hss}}%
	#1}

\usepackage[colorlinks, citecolor=blue,anchorcolor=red,menucolor=red, linkcolor=cyan,filecolor=red,runcolor=red,urlcolor=blue,frenchlinks=red]{hyperref}

 \allowdisplaybreaks

\begin{document}

\title{Recently observed $P_c$ as molecular states and possible mixture of $P_c(4457)$ }

\author{Hao Xu\textsuperscript{1}\footnote{These authors contribute equally to this work.\label{npu}}} \email{xuh2018@nwpu.edu.cn}
\author{Qiang Li\textsuperscript{1\ref{npu}}} \email{liruo@nwpu.edu.cn}
\author{Chao-Hsi Chang$^{2,3,4}$}\email{zhangzx@itp.ac.cn}
\author{Guo-Li Wang$^5$}\email{gl\_wang@hit.edu.cn}

\affiliation{
$^1$School of Physical Science and Technology, Northwestern Polytechnical University, Xi'an 710129, China\\	
$^2${Institute of Theoretical Physics, Chinese Academy of Sciences, Beijing 100190, China}\\
$^3${School of Physical Sciences, University of Chinese Academy of Sciences, 19A Yuquan Road, Beijing 100049, China}\\
$^4${CCAST (World Laboratory), P.O. Box 8730, Beijing 100190, China}\\
$^5${Department of Physics, Hebei University, Baoding 071002, China}
}

\begin{abstract}

{Recently observed spectrum of $P_c$ states exhibits a strong link to $\Sigma_c \bar{D}^{(*)}$ thresholds.
In spite of successful molecular interpretations, we still push forward to wonder whether there exist finer structures. 
Utilizing the effecitve lagrangians respecting heavy quark
symmetry and chiral symmetry, as well as instantaneous Bethe-Salpeter equations, we investigate the $\Sigma_c \bar{D}^{(*)}$ interactions and three $P_c$ states. 
We confirm that $P_c(4312)$ and $P_c(4440)$ are good candidates of $\Sigma_c \bar{D}$ and $\Sigma_c \bar{D}^{*}$ molecules with spin-$\frac12$,
respectively.
 Unlike other molecular calculations, our results indicate $P_c(4457)$ signal {might be} a mixture of 
spin-$\frac32$ and spin-$\frac12$ $\Sigma_c \bar{D}^{*}$ molecules, where the latter one appears to be an excitation of $P_c(4440)$.
Therefore we conclude that, confronting three LHCb $P_c$ signals, there {may} exist not three, but four molecular states.  }

\end{abstract}


\maketitle

\section{introduction}\label{sec1}
The study of exotic states, especially $XYZ$ and pentaquarks, has became a hot topic in recent years. Benefit from upgraded $\tau$-charm 
and $b$ factories such BESIII, LHCb and Belle, large amount of tetraquark candidates, as well as two pentaquark candidates were 
observed. These findings indeed extend our knowledges about non-perturbative QCD. However, the properties and inner structures of the 
states are still in debate, see Refs.~\cite{Chen:2016qju,Guo:2017jvc,Brambilla:2019esw,Liu:2019zoy} for reviews of experimental and 
theoretical status.  

The great progress was made in 2015, when two pentaquark candidates $P_c(4380)$ and $P_c(4450)$ were reported by LHCb collaboration 
\cite{Aaij:2015tga}. With an amplitude analysis, LHCb studied the process $\Lambda_b\to J/\psi K^- p$ and observed them in $J/\psi p$
final states. Both resonances have to be fulfilled with minimal quark content $c\bar{c}uud$, therefore they are good candidates of 
hidden-charm pentaquarks. Just after the discovery, different dynamics were applied to look into their nature: the molecular states of 
$\Sigma_c \bar{D}^{(*)}$ \cite{Chen:2015loa,Chen:2015moa,Karliner:2015ina}, the compact pentaquark structures
\cite{Maiani:2015vwa,Anisovich:2015cia,Wang:2015ava,Lebed:2015tna}, the dynamical effects
\cite{Guo:2015umn,Meissner:2015mza,Liu:2015fea}, and etc. Till now, people have made tremendous effects to clarify their constituents 
and quantum numbers \cite{Chen:2016qju,Liu:2019zoy}. Besides, some predictions have already made before the observations of the $P_c$ 
states \cite{Wu:2010jy,Yang:2011wz,Wang:2011rga,Wu:2012md,Li:2014gra}.

Recently, LHCb \cite{Aaij:2019vzc} re-examined the process $\Lambda_b\to J/\psi K^- p$ with nine-times larger decay samples compared to 
Ref.\,\cite{Aaij:2015tga}, and reveals a more sophisticated structure in $J/\psi p$ invariant mass spectrum than before: $P_c(4450)$ signal
splits into two peaks $P_c(4440)$ and $P_c(4457)$, while a new pentaquark state $P_c(4312)$ shows up in the lower mass
region. The parameters of these $P_c^+$ states are collected in \autoref{Tab-LHCb}.
\begin{table}[h!]
\caption{Summary of the $P_c^+$ properties observed by the LHCb.}\label{Tab-LHCb}
\vspace{0.2em}\centering\setlength{\tabcolsep}{5pt}
\renewcommand{\arraystretch}{1.5}
\begin{tabular}{ cccccccccccc }
\toprule[1.5pt]
State & $M\,[\si{MeV}]$	& $\Gamma\,[\si{MeV}]$ \\
\midrule[1.2pt]
$P_c^+(4312)$	& $4311.9\pm0.7^{+6.8}_{-0.6}$ & $9.8\pm2.7^{+3.7}_{-4.5}$\\
$P_c^+(4440)$	& $4440.3\pm1.3^{+4.1}_{-4.7}$ & $20.6\pm4.9^{+8.7}_{-10.1}$\\
$P_c^+(4457)$	& $4457.3\pm0.6^{+4.1}_{-1.7}$ & $6.4\pm2.0^{+5.7}_{-1.9}$\\
\bottomrule[1.5pt]
\end{tabular}
\end{table}
From \autoref{Tab-LHCb}, we notice that the masses of $P_c(4440)$ and $P_c(4457)$ are slightly below 
$\Sigma_c \bar{D}^{*}$ threshold, while $P_c(4312)$ is quite close to $\Sigma_c \bar{D}$, therefore it is strongly believed that
$\Sigma_c \bar{D}^{(*)}$ interactions are responsible for the enhancements in the $J/\psi p$ invariant spectrum. So far, a number of 
papers have came out to interpret the new results, which carry different opinions such as the molecular states
\cite{Chen:2019asm,Liu:2019tjn,He:2019ify,Xiao:2019aya,Meng:2019ilv,Yamaguchi:2019seo,Valderrama:2019chc,Liu:2019zvb,Huang:2019jlf,Wu:2019adv,Sakai:2019qph,Guo:2019kdc,Xiao:2019mst,Chen:2019bip,Voloshin:2019aut,Guo:2019fdo,Lin:2019qiv,Gutsche:2019mkg,Burns:2019iih,Wang:2019hyc,Du:2019pij,Wang:2019ato}, pentaquarks 
\cite{Weng:2019ynv,Ali:2019npk,Ali:2019clg,Wang:2019got,Giron:2019bcs,Cheng:2019obk,Stancu:2019qga}, 
hadro-charmonium \cite{Eides:2019tgv}, and etc.

{As indicated above, $\Sigma_c \bar{D}^{(*)}$ molecular
interpretations seem to be the most suitable option. Although many papers (such as Refs.~\cite{Chen:2019asm,Chen:2019bip}) have 
confirmed their molecular nature, it is still necessary to examine it from a different approach. Furthermore,
the old $P_c(4450)$ splits into $P_c(4440)$ and $P_c(4457)$, so are there any chances that three $P_c$ states may have 
 finer structures considering the complexity of threshold interaction? }

To answer the question, we will study $\Sigma_c \bar{D}^{(*)}$ interactions and the three $P_c$ states. We first calculate the heavy-hadron interaction amplitudes within the chiral symmetry and heavy quark symmetry\,\cite{Burdman:1992gh,Wise:1992hn,Yan:1992gz,Casalbuoni:1996pg,Cheng:2006dk,Yang:2011wz}, then iterate the obtained interaction kernel into the Bethe-Salpeter equation (BSE) to explore the nature of the $\Sigma_c\bar{D}^{(*)}$ heavy-hadron systems. The Bethe-Salpeter (BS) methods adopted here have 
been successfully applied to investigate the properties of the meson systems, including the mass spectra, hadronic transitions and
weak decays \cite{Chang:2004im,Chang:2010kj,Chang:2005sd,Wang:2011jt,Wang:2013lpa,Wang:2013nya,Li:2016cou,Li:2016efw,Li:2017sww},
as well as the recent $\Xi_{cc}$ study\,\cite{LiQ2020}.
Therefore extending to the meson-baryon molecular systems is quite natural.

Besides, in the BS framework, the relativistic effects and the mixing of the different partial waves can be
automatically involved, in spite of some approximations.
 
It is worth mentioning that, to understand the near threshold phenomenas and resonance formations in a two-hadron system,
a non-perturbative resummation is quite crucial. Such resummation has been considered in chiral dynamics of nucleon-nucleon systems
\cite{Machleidt:2011zz,Epelbaum:2019kcf,Ren:2016jna}
and heavy meson systems \cite{Xu:2017tsr}, as well as the phenomenological studies of molecular states and $XYZ$ exotics (see reviews
\cite{Chen:2016qju,Guo:2017jvc}). A non-perturbative
resummation is to partially summit interactions to all order, which can
be achieved by a proper iterating equation such as the Lippmann–Schwinger equation, the Bethe-Salpeter equation and etc. However, the 
studies mentioned above take simplified or non-relativistic equations. Therefore,
in this article we also want to focus on the BS equation itself to push forward the resummation method. 

In the present work, we will study $P_c(4312)$, $P_c(4440)$ and $P_c(4457)$ states by investigating $\Sigma_c \bar{D}^{(*)}$ 
interactions. First, we adopt the effective Lagrangians with the heavy quark symmetry and chiral symmetry to calculate the transition amplitudes
of the $\Sigma_c \bar{D}^{(*)}$ interactions. Then, we iterate them into the instantaneous Bethe-Salpeter equation, and look for the bound 
state solutions. With careful studies of the meson-baryon interactions, we hope that the natures of three $P_c$ states,
which are astonishingly close to the corresponding thresholds, can be answered. 
One notice that there exists similar works with different approximations: Refs.~\cite{He:2019ify,He:2019rva} studied $P_c(4312)$,
$P_c(4440)$ and $P_c(4457)$ with a quasipotential Bethe-Salpeter equation approach; Ref.~\cite{Ke:2019bkf} stduied one of
 three $P_c$ states $P_c(4312)$ with the Bethe-Salpeter equation.

This work is organized as follows. After introduction, we introduce the effective lagrangians, and present the calculated interaction kernels
(Sec.~\ref{sec2}). In Sec.~\ref{Sec-3}, we exhibit the Bethe-Salpeter equations for $\Sigma_c \bar{D}^{(*)}$ interactions. In Sec.~\ref{Sec-4} numerical results and some discussions are presented.  
Sec.~\ref{Sec-5} denotes to a brief summary and conclusion of this work.

\section{Lagrangians and $\Sigma_c \bar{D}^{(*)}$ interaction kernels} \label{sec2}
To study the $\Sigma_c \bar{D}^{(*)}$ interactions later, we illustrate the corresponding lagrangians first. The interaction between a $S$-wave
heavy-light meson and a light pseudoscalar meson reads \cite{Burdman:1992gh,Wise:1992hn,Yan:1992gz,Casalbuoni:1996pg}
\begin{eqnarray}\label{LagrangianHcpi1}
\mathcal L_{H_cP}&=&
+g\langle \bar{H}_c \slashed u \gamma_5 H_c\rangle.
\end{eqnarray}
In the above, $H_c$ field represents the $(\bar{D},\bar{D}^*)$ doublet in the heavy quark limit
\begin{eqnarray} \label{Hfield}
&& H_c=\left(P^*_{c\mu}\gamma^\mu+iP_c\gamma_5\right)\frac{1-\slashed v}{2},\quad \nonumber \\
&&\bar H_c=\gamma^0 H^\dag_c \gamma^0 = \frac{1-\slashed v}{2}\left(P^{*\dag}_{c\mu} \gamma^\mu+iP^\dag_c \gamma_5\right) ,\nonumber\\
&& P_c=(\bar{D}^0, D^-, D^-_s), \quad P^*_{c\mu}=(\bar{D}^{*0}, D^{*-},D^{*-}_s)_\mu.
\end{eqnarray}
$v=(1,0,0,0)$ stands for the 4-velocity of the $H$ field. The axial vector field $u$ is expressed as
$u_\mu={i\over 2} \{\xi^\dagger, \partial_\mu \xi\}=-\frac{\partial_\mu\phi}{2f}+...$, where $\xi =\exp(i \phi/2f)$, $f$ is the $\pi$ decay constant ( $f_{\textnormal{exp}}=92 \textnormal{ MeV }$), and 
\begin{eqnarray}
\phi&=&\sqrt2\left(\begin{array}{ccc}
\frac{1}{\sqrt{2}}\pi^0+\frac{\eta}{\sqrt{6}}&\pi^+&K^+\\
\pi^-&-\frac{1}{\sqrt{2}}\pi^0+\frac{\eta}{\sqrt{6}}&K^0\\
K^-&\bar{K}^0&-\frac{2\eta}{\sqrt{6}}
\end{array}\right).
\end{eqnarray}

Similarly, the interactions between heavy-light and light-vector (-scalar) mesons read \cite{Casalbuoni:1996pg,Yang:2011wz}
\begin{align}
\mathcal{L}_{HV}&= i\beta\langle \bar{H}_c  v_\mu
V^\mu{H}_c\rangle
+i\lambda\langle \bar{H}_c
\sigma_{\mu\nu}F^{\mu\nu}(V)H_c\rangle, \label{LagrangianHcV}\\
\mathcal{L}_{H\sigma}&=g_s \langle \bar{H}_c\sigma
H_c\rangle, \label{LagrangianHcsigma}
\end{align}
where $F_{\mu\nu}(V)=\partial_\mu V _\nu - \partial_\nu V_\mu +
[V_\mu,{\ } V_\nu]$ and
\begin{eqnarray}
V&=&\frac{ig_V}{\sqrt2}\left(\begin{array}{ccc}
\frac{\rho^{0}}{\sqrt{2}}+\frac{\omega}{\sqrt{2}}&\rho^{+}&K^{*+}\\
\rho^{-}&-\frac{\rho^{0}}{\sqrt{2}}+\frac{\omega}{\sqrt{2}}&K^{*0}\\
K^{*-}&\bar{K}^{*0}&\phi
\end{array}\right).
\end{eqnarray}

The interactions between $S$-wave heavy baryon and light mesons are \cite{Yan:1992gz,Cheng:2006dk,Yang:2011wz}
\begin{eqnarray}
{\cal
	L}_{\mathcal{B}_{\bar{3}}}&=&+
i\beta_B\langle\bar{\mathcal{B}}_{\bar{3}}v^\mu(-V_\mu)
B_{\bar{3}}\rangle
+\ell_B\langle\bar{\mathcal{B}}_{\bar{3}}{\sigma} \mathcal{B}_{\bar{3}}\rangle,\label{B3}\\
{\cal L}_{S}&=&i
\frac{3}{2}g_1\epsilon^{\mu\nu\lambda\kappa}\,v_\kappa\,\langle\bar{\mathcal{S}}_\mu
u_\nu \mathcal{S}_\lambda\rangle
+i\beta_S\langle\bar{\mathcal{S}}_\mu v_\alpha
(-V^\alpha) \mathcal{S}^\mu\rangle \nonumber\\&&+
\lambda_S\langle\bar{\mathcal{S}}_\mu
F^{\mu\nu}(V)\mathcal{S}_\nu\rangle
+\ell_S\langle\bar{\mathcal{S}}_\mu \sigma
\mathcal{S}^\mu\rangle .\label{LagrangianB6}
\end{eqnarray}
Here, $\mathcal{S}_{\mu}^{ab}$ is composed of Dirac spinor
operators
\begin{eqnarray}
\mathcal{S}^{ab}_{\mu}&=&-\sqrt{\frac{1}{3}}(\gamma_{\mu}+v_{\mu})
\gamma^{5}\mathcal{B}_6^{ab}+\mathcal{B}^{*ab}_{6\mu},\\
\bar{\mathcal{S}}^{ab}_{\mu}&=&\sqrt{\frac{1}{3}}\bar{\mathcal{B}}_6^{ab}
\gamma^{5}(\gamma_{\mu}+v_{\mu})+\bar{\mathcal{B}}^{*ab}_{6\mu},
\end{eqnarray}
with 
\begin{eqnarray}
\mathcal{B}_{\bar{3}}&=&\left(\begin{array}{ccc}
0&\Lambda^+_c&\Xi_c^+\\
-\Lambda_c^+&0&\Xi_c^0\\
-\Xi^+_c &-\Xi_c^0&0
\end{array}\right),\\
\mathcal{B}_6&=&\left(\begin{array}{ccc}
\Sigma_c^{++}&\frac{1}{\sqrt{2}}\Sigma^+_c&\frac{1}{\sqrt{2}}\Xi'^+_c\\
\frac{1}{\sqrt{2}}\Sigma^+_c&\Sigma_c^0&\frac{1}{\sqrt{2}}\Xi'^0_c\\
\frac{1}{\sqrt{2}}\Xi'^+_c&\frac{1}{\sqrt{2}}\Xi'^0_c&\Omega^0_c
\end{array}\right).
\end{eqnarray}

We consider two isospin channels ($I=\frac12,\frac32$) of $\Sigma_c \bar{D}^{(*)}$ in our work, which contain eigenstates:
\begin{align}
&\Big|\frac12,\frac12 \Big\rangle=\sqrt{\frac23}\Big|\Sigma_c^{++};\bar{D}^{(*)-}\Big\rangle-\sqrt{\frac13}\Big|\Sigma_c^{+};\bar{D}^{(*)0}\Big\rangle\quad,
\nonumber\\
&\Big|\frac32,\frac12 \Big\rangle=\sqrt{\frac13}\Big|\Sigma_c^{++};\bar{D}^{(*)-}\Big\rangle+\sqrt{\frac23}\Big|\Sigma_c^{+};\bar{D}^{(*)0}\Big\rangle\quad.
\end{align}

With preparations above, we are able to calculate the interaction kernels which will be iterated into the instantaneous
Bethe-Salpeter equation later. These kernels represent tree-level one-meson-exchange diagrams, including $\sigma$, $\pi$, $\eta$, $\rho$ 
and $\omega$ exchanges.

The calculated interaction kernel for $\Sigma_c\bar{D}$ is expressed as
\begin{gather} \label{E-KPB}
K(s_\perp) = F^2(s_{\!\perp}^2) \left(V_1+ V_2 \slash s_{\!\perp}\right),
\end{gather}
where $F(s_{\!\perp}^2)$ denotes the form factor for the interaction vertexes.

In the following, we specify the $g_S$, $\ell_S$, $g$, $g_1$, $\beta$, $\beta_S$, $\lambda$ and $\lambda_S$ 
(in Eqs.~(\ref{LagrangianHcpi1}), (\ref{LagrangianHcV}), (\ref{LagrangianHcsigma}) and (\ref{LagrangianB6}))
to $\sigma_1$, $\sigma_2$,
$\pi_1$, $\pi_2$, $\rho^\up{V}_1$, $\rho^\up{V}_2$, $\rho^\up{T}_1$ and $\rho^\up{T}_2$ respectively, for convenience. For $I=\frac{1}{2}$, potentials $V_1$ and $V_2$ read
\begin{align}\label{E-KVB-Vn}
V_1&= \frac{2\sigma_1 \sigma_2}{E^2_\sigma} + \rho^\up{V}_1 \rho^\up{V}_2 g_V^2 \left(\frac{1}{E^2_\rho} +  \frac{1}{2E^2_\w} \right) , \\
V_2&= \frac{1}{3}\rho^\up{V}_1 \rho^\up{T}_2 g_V^2 \left(\frac{2}{E^2_\rho} - \frac{1}{E^2_\w} \right) .
\end{align}
For $I=\frac{3}{2}$, $V_1$ and $V_2$ are
\begin{align}\label{E-KVB-Vn-I4}
V_1&= \frac{2\sigma_1 \sigma_2}{E^2_\sigma} + \frac{1}{2}\rho^\up{V}_1 \rho^\up{V}_2 g_V^2 \left( \frac{1}{E^2_\w} +\frac{1}{E^2_\rho} \right) , \\
V_2&= -\frac{1}{3}\rho^\up{V}_1 \rho^\up{T}_2 g_V^2 \left(\frac{1}{E^2_\rho} + \frac{1}{E^2_\w} \right) .
\end{align}
In the above, $E_\phi=\sqrt{\vec s^2+m_\phi}$ denotes the energy of the exchanged meson, with $\phi=\sigma$, $\eta$, $\rho$, or $\w$. Notice
that in the $\Sigma_c\bar D$ interaction, only $\sigma$ and $\rho(\w)$ exchanges contribute.

The interaction kernel for $\Sigma_c\bar{D}^*$ is written by
\begin{align} \label{E-KVB}
K^{\alpha\beta}(s_\perp) = &F^2(s_{\!\perp}^2) \bigg[\kappa_1 g^{\alpha\beta}+ \kappa_2 \slash s_{\!\perp} g^{\alpha\beta} + \kappa_3 \slash s_{\!\perp} (\gamma^\alpha s^\beta_{\!\perp} - \gamma^\beta s^\alpha_{\!\perp}) \nonumber\\
& + \kappa_4 \gamma^\alpha \gamma^\beta \bigg].
\end{align}
For the isospin-$\frac{1}{2}$ states, we have potentials
\begin{align}\label{E-KVB-kn}
\kappa_1=&-\frac{2\sigma_1 \sigma_2}{E^2_\sigma} +  \frac{\pi_1\pi_2 \vec s^2}{f^2} \left( \frac{1}{E^2_\pi} - \frac{1}{6 E^2_\eta} \right)  +\rho^\up{V}_1 \rho^\up{V}_2 g_V^2 \bigg(\frac{1}{E^2_\rho}
\nonumber\\
& -  \frac{1}{2E^2_\w} \bigg), \\
\kappa_2=&  \frac{1}{3}\rho^\up{V}_1 \rho^\up{T}_2 g_V^2 \left( \frac{1}{E^2_\w} -\frac{2}{E^2_\rho} \right),\\
\kappa_3=& \frac{\pi_1\pi_2 \vec s^2}{f^2} \left( \frac{1}{E^2_\pi} - \frac{1}{6E^2_\eta} \right)  +\frac{2}{3}\rho^\up{T}_1 \rho^\up{T}_2 g_V^2 \left(\frac{2}{E^2_\rho} -  \frac{1}{E^2_\w} \right), \\
\kappa_4=&-\frac{\pi_1\pi_2 \vec s^2}{f^2} \left( \frac{1}{E^2_\pi} - \frac{1}{6E^2_\eta} \right) .
\end{align}
For the isospin-$\frac{3}{2}$ states,
\begin{align}\label{E-KVB-kn-I4}
\kappa_1=&-\frac{2\sigma_1 \sigma_2}{E^2_\sigma} - \frac{\pi_1\pi_2 \vec s^2}{2f^2} \left( \frac{1}{E^2_\pi} + \frac{1}{3 E^2_\eta} \right) - \frac{1}{2}\rho^\up{V}_1 \rho^\up{V}_2 g_V^2 \bigg(\frac{1}{E^2_\rho} 
\nonumber\\&+ \frac{1}{E^2_\w} \bigg), \\
\kappa_2=& \frac{1}{3}\rho^\up{V}_1 \rho^\up{T}_2 g_V^2 \left( \frac{1}{E^2_\w} +\frac{1}{E^2_\rho} \right),\\
\kappa_3=& -\frac{\pi_1\pi_2 \vec s^2}{2f^2} \left( \frac{1}{E^2_\pi} + \frac{1}{3E^2_\eta} \right) -\frac{2}{3}\rho^\up{T}_1 \rho^\up{T}_2 g_V^2 \left(\frac{1}{E^2_\rho} + \frac{1}{E^2_\w} \right), \\
\kappa_4=&\frac{\pi_1\pi_2 \vec s^2}{2f^2} \left( \frac{1}{E^2_\pi} + \frac{1}{3E^2_\eta} \right) .
\end{align}

\section{the Bethe-Salpeter equations of the $\Sigma_c \bar{D}^{(*)}$ systems}\label{Sec-3}

In this part, we further study the Bethe-Salpeter formalism of the meson-baryon system $\Sigma_c \bar{D}^{(*)}$. 
Considering $J^P=1^-$ or $0^-$ for $\bar{D}^{(*)}$ and $J^P=\frac{1}{2}^+$ for $\Sigma_c$, the corresponding 
Bethe-Salpeter equations and BS wave functions can be obtained.

\subsection{$\Sigma_c(\frac{1}{2}^+) \bar{D}^*(1^-)$ system}
\begin{figure}[htpb]
	\centering
	\includegraphics[width = 0.43\textwidth, angle=0]{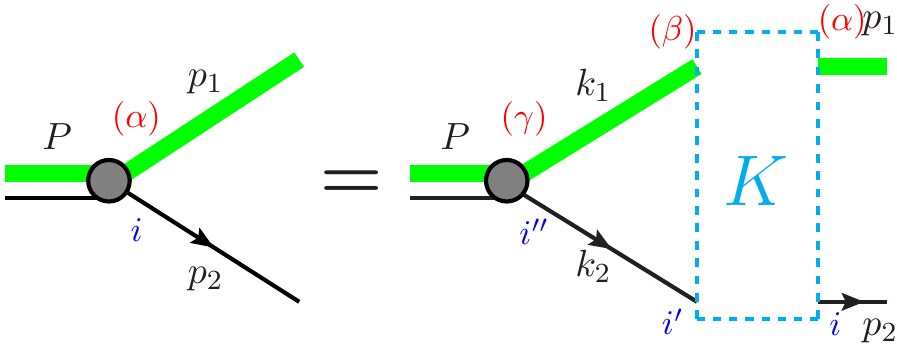}
	\caption{(color online) The Bethe-Salpeter equation of the meson-baryon system. The Greeks~(red) denote the Lorentz indices, while the
	 Romans (blue) represent the Dirac indices. $P,~p_1(k_1)$ and $p_2(k_2)$ stand for the momenta of the pentaquark, meson component, and the baryon component respectively.}\label{Fig-BS-Pc}
\end{figure}

The Bethe-Salpeter equation for a meson-baryon system is schematically depicted in \autoref{Fig-BS-Pc}, which is written by
\begin{align} \label{E-BSE-Pc}
\Gamma^\alpha(P,q,r) =&\int \frac{\up d^4 k}{(2\pi)^4} (-i)K^{\alpha\beta}(P,k_\perp,q_\perp)  [S(k_2) \Gamma^\gamma(P,k,r) \nonumber \\
 &\times D_{\beta\gamma}(k_1)] ,
\end{align}
where $\Gamma(P,q,r)$ denotes the pentaquark (refered as $P_c$ below) vertex carrying total momentum $P$, inner relative momentum $q$, and
spin state $r$. Here we have on-shell condition $P^2=M^2$, with $M$ the $P_c$ mass. The inner relative momenta $q$ and $k$ are defined as
\begin{align}
q=\alpha_2p_1-\alpha_1p_2, \quad k=\alpha_2k_1-\alpha_1k_2,
\end{align}
with $\alpha_{1(2)}\equiv \frac{M_{1(2)}}{M_1+M_2}$. $k_{1(2)}$ and $M_{1(2)}$ are the momentum and mass of the meson component 
(baryon component) respectivly. $S(k_2)=i\frac{1}{\sl k_2 - M_2}$ is the free propagator of the baryon, while the propagator $D^{\alpha\beta}$ (for the $J^P=1^-$ meson) reads
\begin{gather*}
D^{\alpha\beta}(p_1)  =D(p_1)d^{\alpha\beta}(p_{1\perp}),~~~d^{\alpha\beta}(p_{1\perp})= -g^{\alpha\beta}+\frac{p_{1\perp}^\alpha p_{1\perp}^\beta}{M_1^2},
\end{gather*}
where $D(p_1)=i\frac{1}{p_1^2-M_1^2+i\epsilon}$, $p_{1\perp}=p_1-p_1\!\cdot \!v \, v$.

We define a four-dimensional BS wave function and a three-dimensional Salpeter wave function below:
\begin{gather}
\psi_\beta(P,q)=S(p_2) \Gamma^\gamma(P,q,r) D_{\beta\gamma}(p_1) , \label{E-wave-4D}\\
\varphi_\beta(P,q_\perp) \equiv - i\int \frac{\up d q_{_P}}{2\pi} \psi_\beta(P,q),\label{E-wave-3D}
\end{gather}
where $q_{_P}=q\dx v$ and $q_\perp=q-q_{_P} v$.

Performing the contour integral over $q_{_P}$ on both sides of \eref{E-wave-4D} (see appendix\ref{App-1} for details), we obtain the Salpeter equation (SE),
\begin{align} \label{E-SE-0}
\varphi_\alpha(P,q_\perp)=&\frac{1}{2\w_1}\left[ \frac{\Lambda^+ }{M-\w_1-\w_2} +  \frac{\Lambda^-  }{M+\w_1+\w_2}  \right]d_{\alpha\beta}(p_{1\perp})\nonumber\\
&\times \Gamma^\beta(P,q_\perp).
\end{align}
In the above, $\w_i =\sqrt{M_i^2-p_{i\perp}^2}$ with $i=1, 2$ stands for the kinematic energy of the constituent meson or baryon.
We also have the projector operators,
\begin{align}
\Lambda^\pm(p_{2\perp})=\frac{1}{2}\left[1 \pm \hat{H}(p_{2\perp})\right]\gamma^0
\end{align}
with $\hat H=H/\w_2$, and the Dirac Hamilton $ H(p_{2\perp})=(\sl p_{2\perp}+m_2)\gamma^0 $.

We further split $\varphi$ into a positive and a negative energy wave functions:
\begin{align}\label{SalpeterWaveFunctionPN}
\varphi_\beta=\varphi_\beta^++\varphi_\beta^-, \quad
\varphi^{\pm}_\beta(P,q_\perp)=\Lambda^\pm \gamma^0 \varphi_\beta.
\end{align}
Notice in the weak binding condition $M\sim (\w_1+\w_2)$, we have $\varphi^+\gg \varphi^-$, i.e. the positive energy wave function $\varphi^+(q_\perp)$ dominates. Combined with \eref{SalpeterWaveFunctionPN}, the SE of $\Sigma_c \bar{D}^*$ system can be further
simplified to a ``Shr\"odinger-like" equation:
\begin{gather} \label{E-VS-SE}
M \varphi_\alpha =  (\w_1+\w_2) \hat H(p_{2\perp})\varphi_\alpha +\frac{ d_{\alpha\beta}\gamma^0 \Gamma^\beta(q_\perp) }{2\w_1},
\end{gather}
with $\Gamma^\beta(q_\perp)$ the integral of Salpeter wave function and the kernel,
\begin{align} \label{E-Gamma-3D}
\Gamma^\beta(q_\perp) =\int \frac{\up d^3 k_\perp}{(2\pi)^3}  K^{\beta\gamma}(s_\perp)  \varphi_\gamma(P,k_\perp) .
\end{align}
Notice that the interaction kernel $K(s_\perp)$ is assumed to be instantaneous, thus has no dependence on the time component of the momentum transfer $s=(k-q)$. 

Let us focus on the three-dimensional BSE (\ref{E-VS-SE}). We can see that, the Salpeter wave function $\varphi_\alpha(q_\perp)$ in
\eref{E-SE-0} is just transformed to an integral-type eigenvalue
equation. In \eref{E-VS-SE}, the first term of the right side,
which is determined by Dirac Hamiltonian $H$, stands for the kinetic energy. The second term contains the interaction kernel $K$, therefore represents the potential energy.

In general, the normalization of a BS wave function is expressed as
\begin{align} \label{BSNormalization}
&-i\int \int \frac{\up{d}^4 q}{(2\pi)^4}\frac{\up{d}^4 k}{(2\pi)^4} \bar{\psi}_\alpha(P,q,\bar r) \frac{\partial}{\partial P^0} I^{\alpha\beta}(P,k,q) \psi_\beta(P,k,r)   \nonumber\\
&=2M \delta_{r\bar r} ,
\end{align}
where
\begin{align}
I^{\alpha\beta}(P,q,k)=&(2\pi)^2 \delta^4(k-q) S^{-1}(p_2) D^{-1  \alpha\beta}(p_1) \nonumber\\
&+ iK^{\alpha\beta}(P,k,q).
\end{align}
In the above, we have the inverse of the vector propagator \[ D^{-1}_{\alpha\beta}(p_1)=\vartheta_{\alpha\beta}D^{-1}(p_1),\]
with $\vartheta^{\alpha\beta}=-g^{\alpha\beta}+\frac{p^\alpha_{1\perp} p^\beta_{1\perp}}{\w_1^2}$ as well as the identity $\vartheta^{\alpha\beta}d_{\beta\gamma}=\delta^{\alpha}_{\gamma}$.

As mentioned before, the interaction kernel is assumed to be no dependence on $P^0$ and $q_P$, namely, $K^{\alpha\beta}(P,k,q)\simeq K^{\alpha\beta}(s_\perp)$, therefore the normalization would only involve the term related to two inversed propagators.
After some deduction, \eref{BSNormalization} can be further simplified to
\begin{align}\label{E-Norm-D1}
\int \frac{\up{d}^3 q_\perp}{(2\pi)^3} 2\w_1  \vartheta^{\alpha\beta}  \bar\varphi_\alpha(q_\perp,\bar r) \gamma^0   \varphi_\beta(q_\perp,r)=2M\delta_{r\bar r},
\end{align}
which is just the normalization condition of Salpeter equation (\ref{E-wave-3D}).

\subsection{$\Sigma_c(\frac{1}{2}^+) \bar{D}(0^-)$ system}
Similarly, the Bethe-Salpeter equation for $\Sigma_c \bar{D}$ system reads
\begin{gather} \label{E-BSE-Pc-1}
\Gamma(P,q,r) =\int \frac{\up d^4 k}{(2\pi)^4} (-i)K(P,k_\perp,q_\perp)  [S(k_2) \Gamma(P,k,r) D(k_1)] ,
\end{gather}
where $\Gamma(P,q,r)$ denotes $P_c$ vertex. The BS wave function $\psi$ and related Salpeter wave function $\varphi$ are also 
defined as
\begin{gather}
\psi(P,q)=S(k_2) \Gamma(P,q,r) D(q_1) , \label{E-wave1-4D}\\
\varphi(P,q_\perp) \equiv - i\int \frac{\up d q_P }{2\pi} \psi(P,q)
\end{gather}
Performing the contour integral on $q_P$ over both sides of \eref{E-wave1-4D}, we obtain the Salpeter equation
\begin{gather} \label{E-VS1-SE}
M \varphi(q) =  (\w_1+\w_2) \hat H(p_{2\perp})\varphi +\frac{ \gamma^0 \Gamma(q) }{2\w_1}, 
\end{gather}
with
\begin{align} \label{E-Gamma1-3D}
\Gamma(q_\perp) =\int \frac{\up d^3 k_\perp}{(2\pi)^3}  K(s_\perp)  \varphi(P,k_\perp) .
\end{align}

Applying the same strategy above, we obtain the normalization of $\varphi$:
\begin{align}\label{E-Norm1-D1}
\int \frac{\up{d}^3 q_\perp}{(2\pi)^3} 2\w_1  \bar\varphi(q_\perp,\bar r) \gamma^0 \varphi(q_\perp,r)=2M\delta_{r\bar r}.
\end{align}

\subsection{The constructions of the Salpeter wave functions and further reductions}

We first turn to $\Sigma_c(\frac{1}{2}^+) \bar{D}(0^-)$ system. Accounting the spin-parity and the Lorentz structures, the Salpeter wave function ($J^P=\frac{1}{2}^-$) can be constructed as
\begin{gather} \label{E-wave1-1-2N}
\varphi(P,q_\perp,r) =A(q_\perp)\gamma^5 u(P,r)=\left (f_1+f_2 \frac{\sl q_\perp}{q} \right) \gamma^5 u(P,r) ,
\end{gather}
where $f_{1(2)}(\vabs{q})$ only depends on $\vabs{q}$. 

It is worth mention that, the wave function above can be rewritten  in terms of the spherical harmonics $Y_l^m$:
\begin{align} \label{E-wave0-Ylm}
\varphi(P,q_\perp,r)=& C_0 \left[f_1 Y_0^0  + C_1 f_2 \left( Y_1^{+1}\gamma^- + Y_1^{-1} \gamma^+ -Y_1^0 \gamma^3\right)   \right]
\nonumber\\ &\times \gamma^5 u(P,r),
\end{align}
where $C_0=2 \sqrt{\pi}$ and $C_1=\frac{1}{\sqrt{3}}$; $\gamma^{\pm} = \mp\frac{1}{\sqrt{2}} (\gamma^1\pm i \gamma^2)$. Therefore it
is quite obvious that $f_1$ and $f_2$ represent $S$- and $P$-wave components, respectively.

By inserting the wave function into \eref{E-Norm1-D1}, we obtain the normalization
\begin{align}\label{E-Norm1-D12}
\int \frac{\up{d}^3 q_\perp}{(2\pi)^3} 2\w_1 \left (f_1^2+f_2^2 \right )=1.
\end{align}

The $\frac{1}{2}^-$ Salpeter wave function composited of $\Sigma_c(\frac{1}{2}^+) \bar{D}^*(1^-)$ can be written as
\begin{gather} \label{E-wave-1-2N}
\varphi_{\alpha}(P,q_\perp,r) =  A_\alpha(q_\perp) u(P,r) ,
\end{gather}
with
\[
A_\alpha= \left( g_1+ g_2 \frac{\sl q_\perp}{q}  \right)(\gamma_\alpha -v_\alpha)  + \left( g_3+ g_4 \frac{\sl q_\perp}{q} \right)\hat q_{\perp\alpha},
\]
where $\hat q_{\perp\alpha}= \frac{q_{\perp\alpha}}{ |\vec q\,|}$, $u(P,r)$ represents the Dirac spinor carrying momentum $P$
and spin state $r$. Notice that the radial wave function $g_i(\vabs{q})~(i=1,\cdots,4)$ only depends on $\vabs{q}$. It is clear that $g_1$
corresponds to $S$ wave, $g_{2(3)}$ belongs to $P$ wave, and $g_4$ contributes both to $S$ and $D$ partial waves
(see Ref.~\cite{LiQ2020} for a further reading about different partial waves in terms of the spherical harmonics $Y_l^m$). 

Inserting \eref{E-wave-1-2N} into \eref{E-Norm-D1}, we obtain following normalization condition
\begin{align}
&\int \frac{\up{d}^3 q_\perp}{(2\pi)^3} 2\w_1  \left[
3c_3 \left(g_1^2+g_2^2 \right) + c_1 \left(g_3^2+g_4^2-2g_1g_4+2g_2g_3\right)\right]
\nonumber\\ 
&=1,
\end{align}
where $c=-{q^2}/{\w_1^2}$, $c_1 = 1+c$, $c_3=1+c/3$.

For the $\frac{3}{2}^-$ state with $\Sigma_c(\frac{1}{2}^+) \bar{D}^*(1^-)$, the Salpeter wave function is written by
\begin{align} \label{E-wave-3-2N}
\varphi_{\alpha}(P,q_\perp,r) 
&= A_{\alpha\beta} \gamma^5 u^\beta (P,r),
\end{align}
where
\begin{align}
A_{\alpha\beta}(q_\perp) \equiv &\left( h_1+ h_2  \frac{\sl q_\perp}{q} \right ) g_{\alpha\beta}+ \left( h_3+ h_4 \frac{\sl q_\perp}{q} \right)(\gamma_\alpha +v_\alpha) \hat q_{\perp\beta} \nonumber\\
& + \left(h_5+h_6 \frac{\sl q_\perp}{q} \right) \hat q_{\perp\alpha} \hat q_{\perp\beta}.\nonumber
\end{align}
In \eref{E-wave-3-2N}, $ u^\beta (P,r)$ is a Rarita-Schwinger spinor with polarization $r=\pm\frac{3}{2},~\pm\frac{1}{2}$. Note that the 
constructed Salpeter wave functions (\ref{E-wave-1-2N}), (\ref{E-wave-3-2N}) fulfills the condition $P^\alpha \varphi_\alpha=0$.

The normalization of \eref{E-wave-3-2N} is calculated to be
\begin{align}
&\int \frac{\up{d}^3 q_\perp}{(2\pi)^3} 2\w_1  \bigg[
c_3 \sum_{i=1}^4 h_i^2 +\frac{1}{3}c_1\bigg(h_5^2+h_6^2+2h_4h_5 \nonumber\\
&-2 h_1h_5-2h_2h_6-2h_3h_6\bigg)+\frac{2}{3} c \left(h_2h_3-h_1 h_4\right)
\bigg]=1,
\end{align}
where the following completeness relation \cite{Behrends1957} has been used:
\begin{align}\notag
&\sum_r u^\alpha(P,r) \bar u^\beta(P,r)\nonumber\\
&=(\sl P+M)\left[-g^{\alpha\beta}+\frac{1}{3}\gamma^\alpha\gamma^\beta - \frac{P^\alpha\gamma^\beta-P^\beta\gamma^\alpha}{3M}+\frac{2P^\alpha P^\beta}{3M^2} \right].
\end{align}

With above preparations, we are ready to transform Eqs.~(\ref{E-VS-SE}) and (\ref{E-VS1-SE}) into a set of coupled eigenvalue equations. 
For example, inserting the Salpeter wave function (\ref{E-wave1-1-2N}) into \eref{E-VS1-SE}, we obtain
\begin{align} \label{E-Eigen1-1-2}
&\left[M-(\w_1+\w_2) \hat H(p_{2\perp})\right] A(q_\perp) \gamma^5 u(P,r) \nonumber\\
&= \frac{1}{2\w_1}  \gamma^0 \int \frac{\up{d}^3 k_\perp}{(2\pi)^3} K(s_{\!\perp}) A \gamma^5 u(P,r).
\end{align}
After 
eliminating the spinor as well as projecting out the radial wave function, we obtain two coupled eigenvalue equations for
$\Sigma_c \bar{D}$ system
\begin{align}
Mf_1(\vabs{q}) =&R_{11} f_1(\vabs{q}) + R_{12} f_2(\vabs{q}) -\frac{1}{ 2\w_1}\int \frac{\up{d}^3 \vec k}{(2\pi)^3} \bigg[V_1 f_1(\vabs{k})
\nonumber\\
&+\frac{\vec s\cdot \vec k}{k}V_2 f_2(\vabs{k})\bigg],\\
Mf_2(\vabs{q}) =&R_{21} f_1(\vabs{q}) + R_{22} f_2(\vabs{q}) -\frac{1}{ 2\w_1}\int \frac{\up{d}^3 \vec k}{(2\pi)^3} \bigg[\frac{\vec s\cdot \vec q}{q}
\nonumber\\
&\times V_2 f_1(\vabs{k})-\frac{\vec k\cdot \vec q}{kq} V_1 f_2(\vabs{k})\bigg],
\end{align}
where $R_{22}=-R_{11}={m_2(\w_1+\w_2)}/{\w_2}$, $R_{12}=R_{21}={q(\w_1+\w_2)}/{\w_2}$. Here, the eigenvalue $M$ is just $P_c$ mass.
Solving these equations numerically, the corresponding mass spectra and wave functions can be obtained.

By inserting Eqs.~(\ref{E-wave-1-2N}) and (\ref{E-wave-3-2N})  into \eref{E-VS-SE}, we can also work out the coupled eigenvalue equations for $\Sigma_c \bar{D}^*$ system with $J^P=\frac{1}{2}^-$,~$\frac{3}{2}^-$. 

\section{Numerical results and Decoding three $P_c$ states} \label{Sec-4}
In order to perform the numerical calculations, we first specify the values of the parameters used in this work
\cite{Yang:2011wz,Wang:2019nwt,Chen:2015loa,Chen:2019asm,He:2019ify}:
\begin{align}
\sigma_1&=0.76,&~~\pi_1&=0.59,&~~\rho^\up{V}_1&=0.9,&~~\rho^\up{T}_1&=0.56\,\si{GeV}^{-1}, &\nonumber\\
\sigma_2&=6.2,&~~\pi_2&=0.94,&~~\rho^\up{V}_2&=1.74,&~~\rho^\up{T}_2&=3.31\,\si{GeV}^{-1},& \nonumber\\
g_V&=5.9.
\end{align}

We apply a monopole form factor in our work:
\begin{gather}
F(s_{\!\perp}^2)= \frac{\Lambda^2}{-s_{\!\tiny{\perp}}^2+\Lambda^2},
\end{gather}
where $\Lambda$ is a parameter that characterizes the shape of the form factor, and usually set to the energy scale of the meson
exchange. In our case, $\Lambda=0.12\,\si{GeV}$ for $\Sigma_c \bar{D}^{*}$ and $0.18\,\si{GeV}$ for $\Sigma_c \bar{D}$. Notice that in the limit  $s^2\to 0$, the heavy hadrons are treated as free and point-like particles, therefore the form
factor is normalized to 1 at $s^2\to 0$. 

\begin{figure}[htpb]
	\vspace{0.5em}
	\centering
	\subfigure[$V_i\cdot F^2~(i=1,2)$ ]{\includegraphics[width=0.34\textwidth,angle=0]{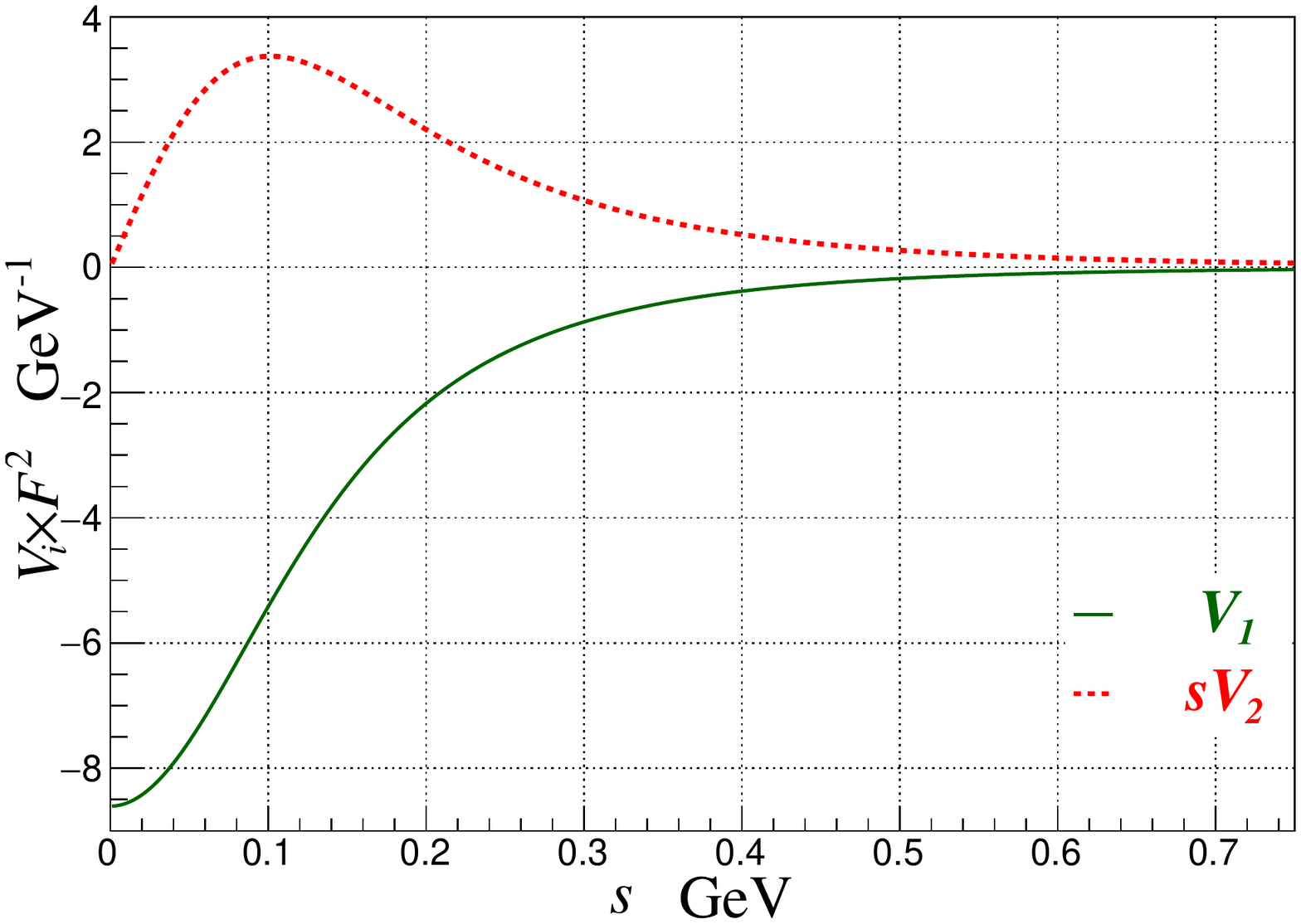} \label{Fig-V2}}
	\subfigure[$\kappa_i\cdot F^2~(i=1,\cdots,4)$ ]{\includegraphics[width=0.34\textwidth,angle=0]{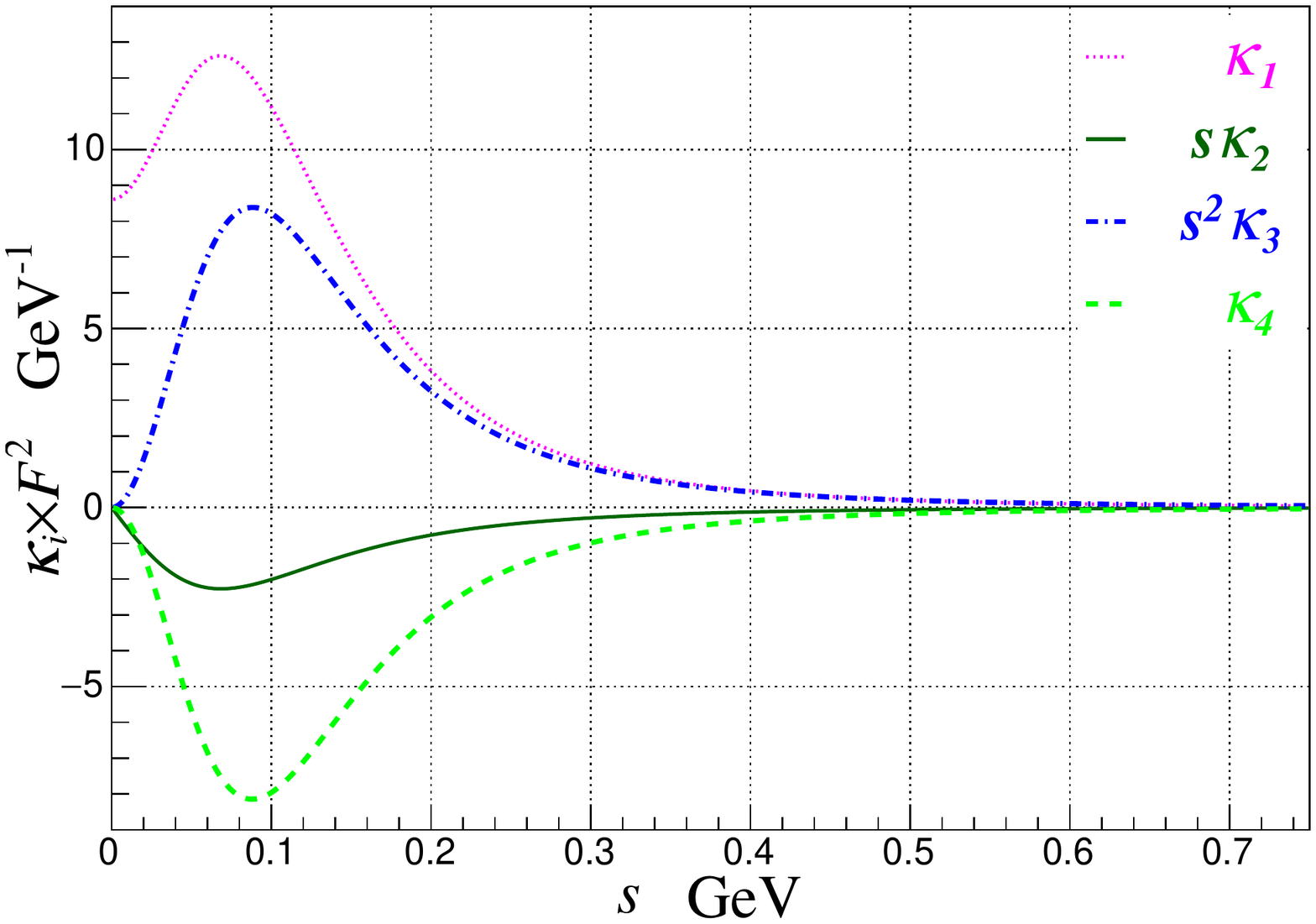} \label{Fig-V4}}
	\caption{The isospin-$\frac{1}{2}$ potentials $V_i\!\cdot\! F^2~(i=1,2)$ and $\kappa_n\cdot F^2~(n=1,\cdots,4)$ for $\Sigma \bar{D}$ and $\Sigma_c\bar{D}^*$, respectively.}\label{Fig-Vn}
	\vspace{0.5em}
\end{figure}

First, we illustrate the results of the potential $V_i$ and $\kappa_i$ with $I=\frac{1}{2}$ appearing in Eqs.~(\ref{E-KPB})
and (\ref{E-KVB}).
Their $s$ (transferred momentum) dependences are depicted in \autoref{Fig-Vn}. 

\begin{table}[h!]
	\caption{Calculated mass spectrum and binding energy $\Delta E$ (in MeV) of $\Sigma_c \bar{D}^{(*)}$ system with $I=\frac{1}{2}$, as
		well as categorized $P_c$.}\label{Tab-Mass}
	\vspace{0.2em}\centering
	\setlength{\tabcolsep}{5pt}
\renewcommand{\arraystretch}{1.5}
	\begin{tabular}{ c|ccccccccccc }
		\toprule[1.5pt]
		$\Sigma_c \bar{D}^{(*)}\big(J^P\big)$ 	& $M$	 	&	$\Delta E$	& $P_c$	&  \\
		\midrule[1.2pt]
		$\Sigma_c \bar{D}\big(\frac12^-\big)$	&$4313^{-2}_{+2}$	        &$-5^{+2}_{-2}$	                &$P_c(4312)$  \\
		$\Sigma_c \bar{D}^{*}\big(\frac12^-\big)$	&$ 4440^{-5}_{+ 4}$	      &$ -20^{+5}_{-4}$                 &$P_c(4440)$ \\
		$\Sigma_c \bar{D}^{*}\big(\frac32^-\big)$	&$4457^{-2}_{+ 1}$     &$-3^{+2}_{-1}$                 &$P_c(4457)$ \\
		$\Sigma_c \bar{D}^{*}\big(\frac12^-\big)$  &$4456^{-1}_{+ 1}$      &$-4^{+2}_{-1}$                &$P_c(4457)$ \\
		\bottomrule[1.5pt]
	\end{tabular}
\end{table}

After solving the relevant eigenvalue equations, we find bound state solutions for $I=\frac12$ $\Sigma_c \bar{D}^{(*)}$ systems.
The obtained mass spectra and corresponding binding energy are listed in \autoref{Tab-Mass}. To see the sensitivities of the 
calculations, we also vary $\Lambda$ by $\pm5\%$ as uncertainties. 

 For $\Sigma_c \bar{D}$, 
with the reasonable parameter $\Lambda$ the mass of experimental $P_c(4312)$ can be well reproduced, i.e., we obtain {the meson-baryon bound state with mass} $4.313$\,GeV.
 Therefore $P_c(4312)$ is a good candidate of $I=\frac12$ $\Sigma_c \bar{D}$ molecular state carrying $J^P=\frac12^-$.
 
For $\Sigma_c \bar{D}^{*}$ system, we obtain two bound states: spin-$\frac{1}{2}$ with mass $4.440$ GeV and
spin-$\frac{3}{2}$ with $4.457$ GeV.
These two are consistent well with the experimental $P_c(4440)$ and $P_c(4457)$ respectively. We conclude $P_c(4440)$ can be treated
as $I=\frac12$ $\Sigma_c \bar{D}^*$ molecular state carrying $J^P=\frac12^-$, while $P_c(4457)$ is $I=\frac12$ $\Sigma_c \bar{D}^*$ molecular state carrying $J^P=\frac32^-$.

However, it is not the end of our story. Differed from other molecular calculations, our approaches indicate an additional state in
the $I=\frac12$ $\Sigma_c \bar{D}^{*}$ channel. This $P_c$ state is an excitation of $P_c(4440)$ which carries
$J^P=\frac12^-$. The most interesting thing is, it has mass $M=4.456\,\si{GeV}$, which is located right at $P_c(4457)$ mass region. Therefore we speculate that $P_c(4457)$ signal discovered by LHCb may contain two overlapped signals: spin-$\frac12$ one
and spin-$\frac32$ one. 

We now refer the spin-$\frac12$ signal as $P_c^\prime(4457)$. In a word, we totally determine four $\Sigma_c \bar{D}^{(*)}$ molecular
states: $P_c(4312)$, $P_c(4440)$, $P_c(4457)$ and $P_c^\prime(4457)$, where the last two are mixed as observed signal in $J/\psi p$
mass spectrum \cite{Aaij:2019vzc}. 

\begin{figure*}[htbp]
\vspace{0.5em}
\centering
\subfigure[$P_c(4312)$ ]{\includegraphics[width=0.34\textwidth,angle=0]{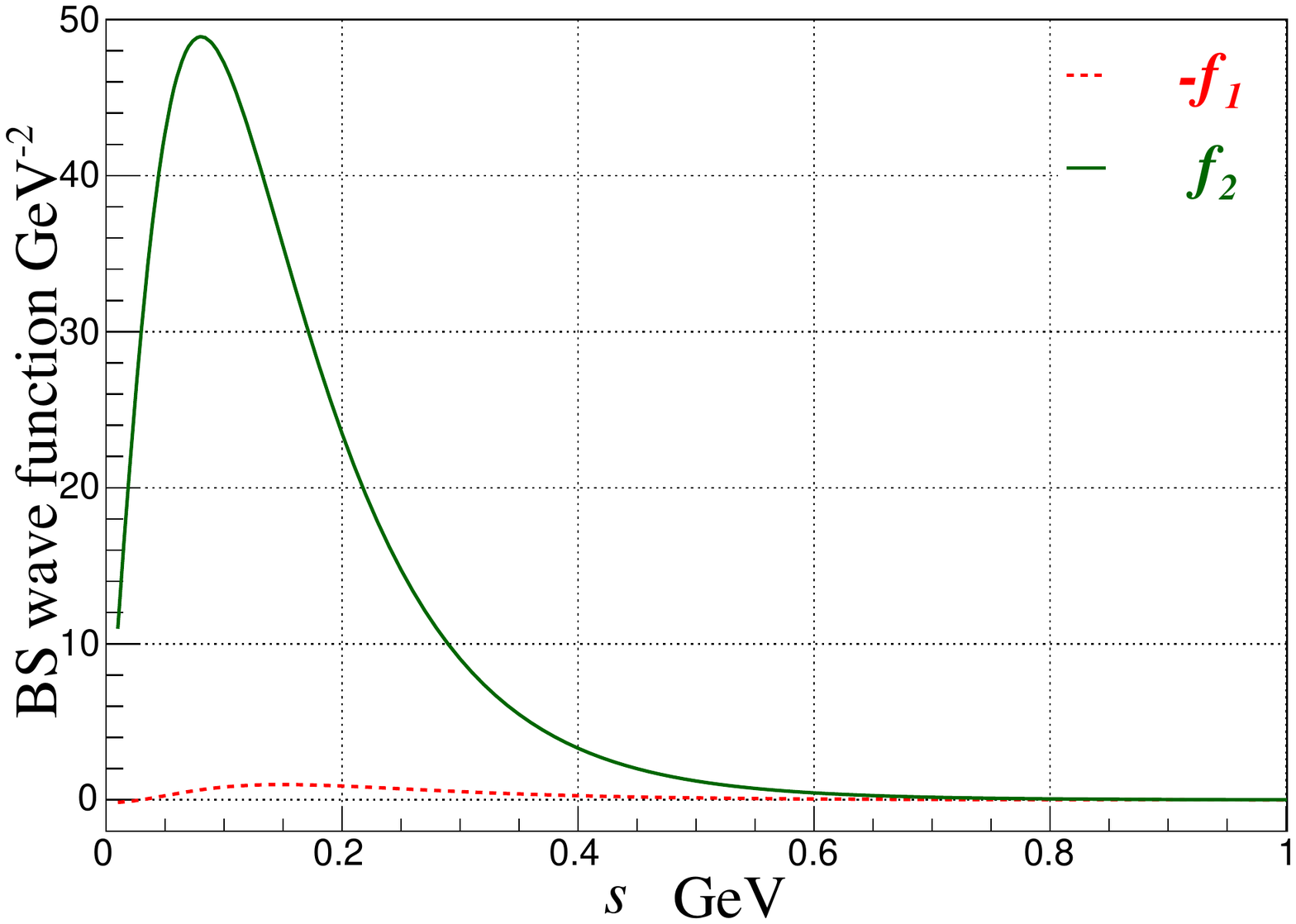} \label{Fig-wave-1}}~~
\subfigure[$P_c(4440)$ ]{\includegraphics[width=0.34\textwidth,angle=0]{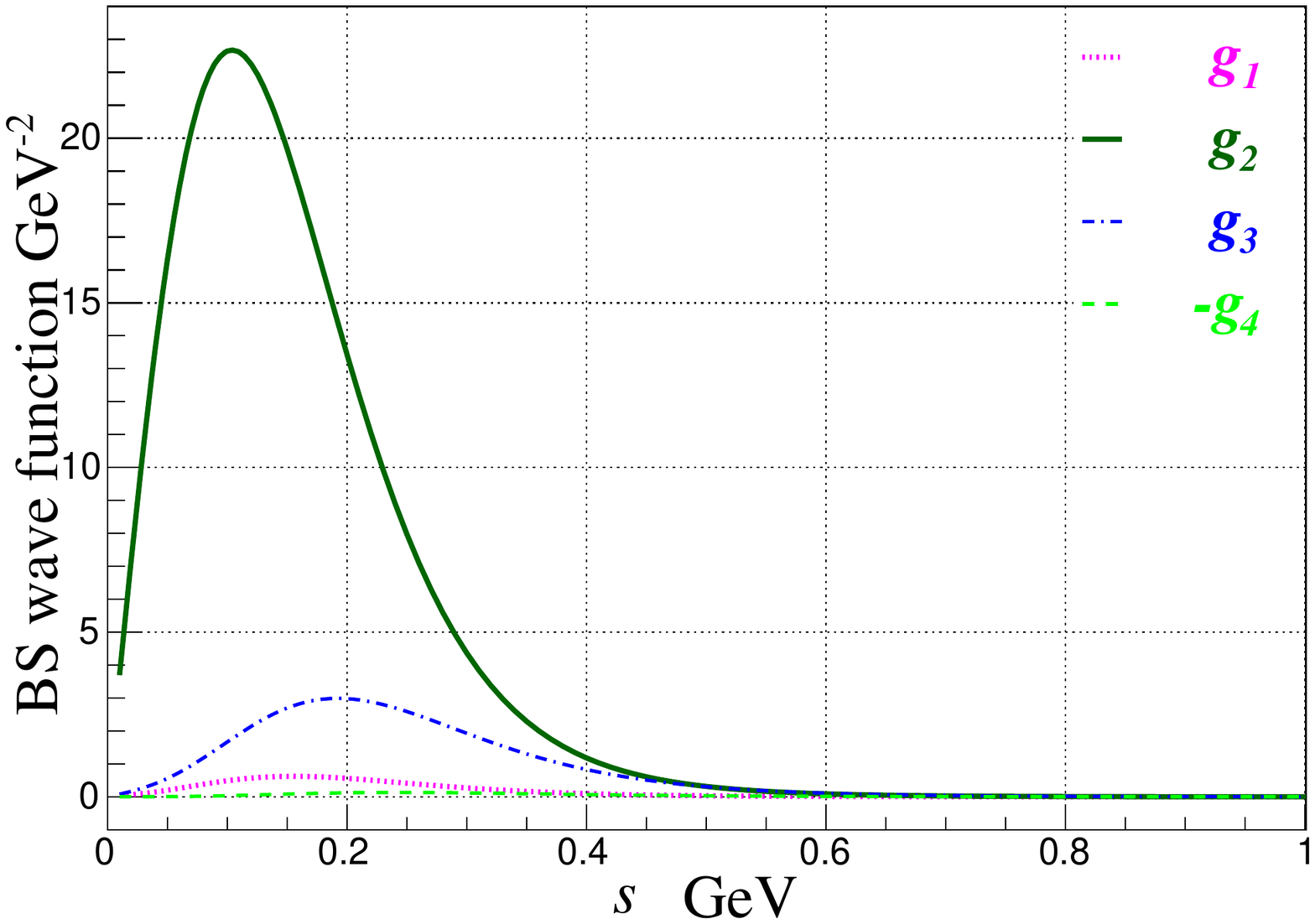} \label{Fig-wave-2}}
\subfigure[$P_c(4457)$ ]{\includegraphics[width=0.34\textwidth,angle=0]{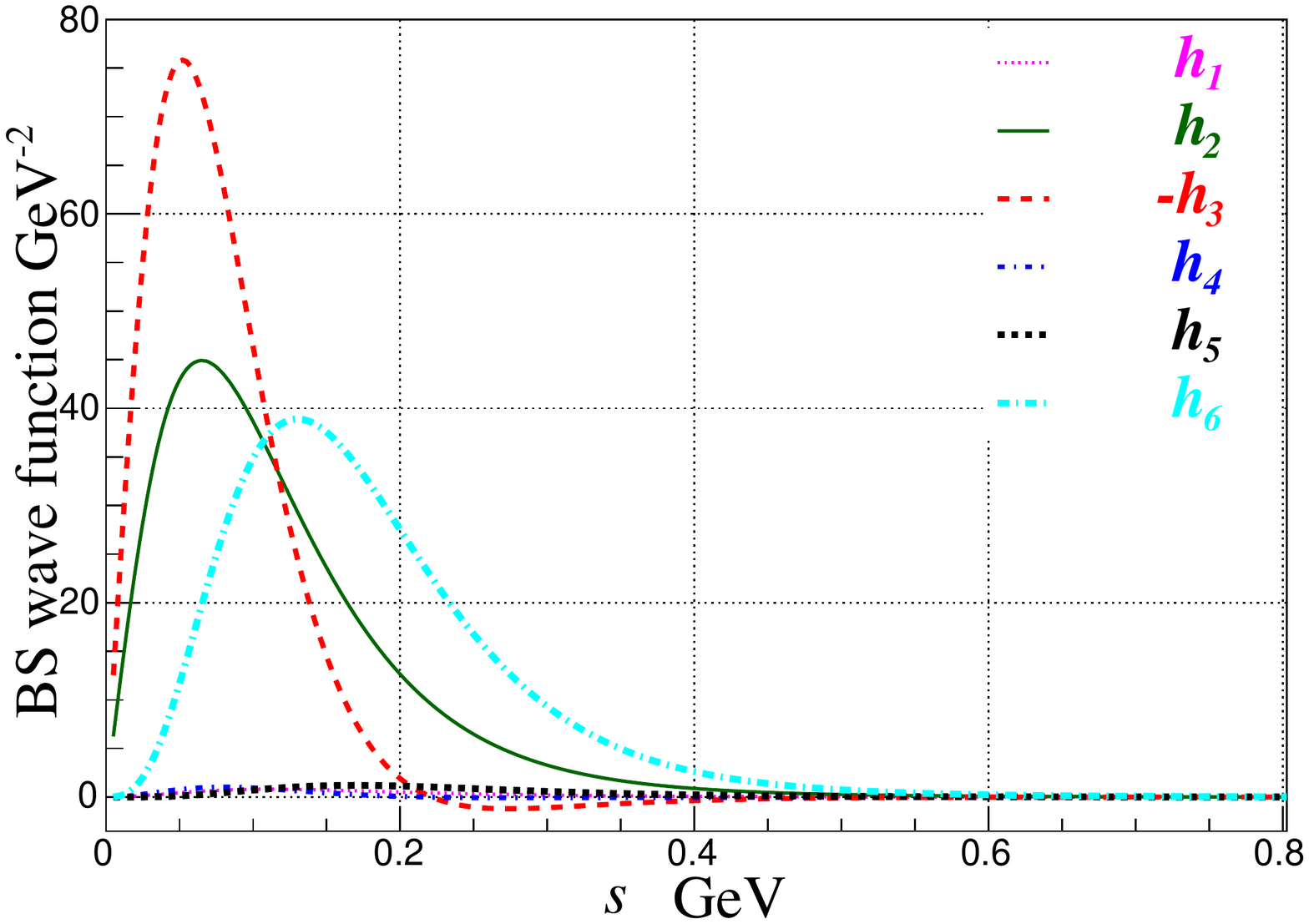} \label{Fig-wave-3}}~~
\subfigure[$P_c^\prime(4457)$ ]{\includegraphics[width=0.34\textwidth,angle=0]{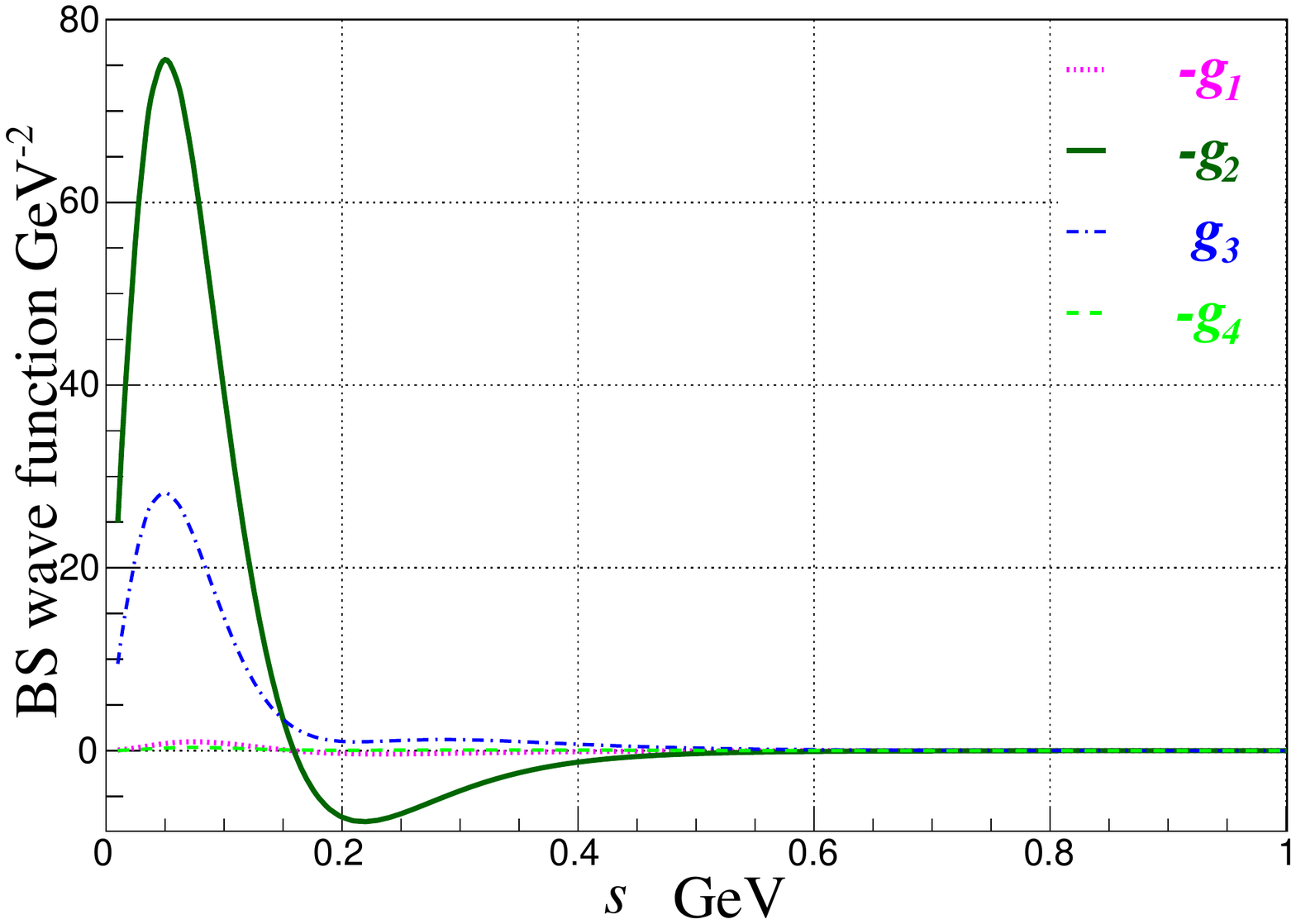} \label{Fig-wave-2B}}
\caption{The BS radial wave functions of (a) $P_c(4312)$, (b) $P_c(4440)$, (c) $P_c(4457)$ and (d) $P_c^\prime(4457)$.}\label{Fig-wave}
\vspace{0.5em}
\end{figure*}

The BS wave functions of $P_c(4312)$, $P_c(4440)$, $P_c(4457)$ and $P_c^\prime(4457)$ are displayed in \autoref{Fig-wave}. We can see
that, $f_2$ is dominant in $P_c(4312)$'s wave function, while $g_2$ is quite prominent in the wave functions of $P_c(4440)$ and
$P_c^\prime(4457)$. In general, we observe that the wave functions of the $P_c$ states are mixtures of $S$, $P$, $D$ waves and even radial
exited components. Notice that in our framework, there only exists limited number (four in our case) of bound states.

$P_c^\prime(4457)$ predicted in our work is mainly a first radial excitation, which means it has a similar property comparing
to $P_c(4440)$. Furthermore the mass gap between $P_c(4440)$ and $P_c^\prime(4457)$ are just $\sim17$ MeV, we believe their widths are
quite close. However, as a radial excitation, we prefer a smaller production ratio of $P_c^\prime(4457)$, therefore it is
reasonable that LHCb can describe 4457 MeV peak now without additional $P_c^\prime(4457)$.

Indeed, with the limited informations in \cite{Aaij:2019vzc} we can not trace $P_c^\prime(4457)$ for now. 
 We expect that LHCb can further investigate
quantum numbers and more decay channels, as well as add $P_c^\prime(4457)$ in their amplitude analysis, to justify our predictions.
For example, LHCb can include only spin-$\frac12$ or both spin-$\frac32$ and spin-$\frac12$ at 4457 Mev for comparing. 
On the other hand, for providing more useful and specific informations, we will theoretically investigate the decay properties and production mechanism in future work.

In addition, we did not find any $I=\frac32$ partners of the $\Sigma_c \bar{D}^{(*)}$ systems.

\section{Summary} \label{Sec-5}
{Recently, LHCb re-examined the process $\Lambda_b\to J/\psi K^- p$, and discovered three $P_c$ pentaquarks: $P_c(4312)$, $P_c(4440)$ and $P_c(4457)$ \cite{Aaij:2019vzc}. Although their molecular nature has been confirmed by many papers, we still think there may emerge 
interesting structures among three $P_c$ signals considering the experience that old $P_c(4450)$ splits into $P_c(4440)$ and $P_c(4457)$.
}

Benefited from the effective lagrangians respecting the chiral and heavy quark symmetry, as well as the instantaneous
Bethe-Salpeter equation, we investigate $\Sigma_c \bar D^{(*)}$ interactions and observed $P_c$ signals. First, we calculate 
the $\Sigma_c \bar D^{(*)}$ interaction amplitudes according to the effective lagrangians, and study the behaviors of the 
potentials (\autoref{Fig-Vn}). Then, we study the BS formalism of the $\Sigma_c \bar D^{(*)}$ system, iterate former interactions
into the BS equations. Finally, we obtain molecular solutions as well as their BS wave functions (\autoref{Fig-wave}).

Our calculations show that, $P_c(4312)$ can be treated as $I=\frac12^-$ $\Sigma_c \bar{D}$ molecular state with $J^P=\frac12^-$,
while $P_c(4440)$ is a good candidate of $I=\frac12$ $\Sigma_c \bar{D}^{*}$ molecule, which also carries $J^P=\frac12^-$.

Differed from other molecular calculations, our work indeed indicate not one, but two bound states in $P_c(4457)$ {mass} region: one is 
$\Sigma_c \bar{D}^{*}$ with $J^P=\frac32^-$, another is $\Sigma_c \bar{D}^{*}$ carrying $J^P=\frac12^-$, which is just an excitation
of $P_c(4440)$. Therefore we conclude that $P_c(4457)$ signal discovered by LHCb {might} be a mixture of $J^P=\frac12^-$ and
$\frac32^-$ states.

In a word, we totally determine four molecular states $P_c(4312)$, $P_c(4440)$, $P_c(4457)$ and $P_c^\prime(4457)$,
which matches to three LHCb signals. We speculate the existence of the additional excitation ($P_c^\prime(4457)$) is necessary, because
the relatively large excitation space of ground $P_c(4440)$. Also, $P_c^\prime(4457)$ has very similar decay property with $P_c(4440)$.

Moreover, we did not support any existences of corresponding isospin-$\frac32$ molecular solutions in our calculations.

We expect LHCb can perform amplitude analysis with more data samples, as well as search for other decay channels, to separate
two states in 4457 MeV signal region. We hope our conclusions can be testified in the future.

\section*{Acknowledgments}
We thank Xiang Liu for helpful discussions. This work is supported by the
Fundamental Research Funds for the Central Universities, under Grant No.\,31020180QD118 and No.\,310201911QD054.  
This work is supported in part by the National Natural Science Foundation of China (NSFC) under Grant Nos.\, 11575048, 11745006, 11535002, 11675239,  and 11821505.

\appendix

\section{Some expressions for derivations of the Salpeter equation}\label{App-1}
To perform the contour integral, we 
rewrite the propagators: 
\begin{gather}
S(p_2)=i\frac{1}{\sl{p}_2-m_2}=-i\left(\frac{\Lambda^+}{q_P-\zeta_2^+-i\epsilon }+\frac{\Lambda^-}{q_P-\zeta_2^-+i\epsilon}\right),\\
D(p_1)=i\frac{1}{p_1^2-m_1^2}=i\frac{1}{2\w_1}\left(\frac{1}{q_P-\zeta_1^++i\epsilon }-\frac{1}{q_P-\zeta_1^--i\epsilon}\right),
\end{gather}
where
$\zeta_2^\pm=\alpha_2 M\mp\w_2$,
$\zeta_1^+=-\alpha_1 M \pm \w_1$.

Inserting above expressions into \eref{E-wave-4D}, then performing the contour integral over $q_P$, we obtain the
three-dimensional Salpeter equation (\ref{E-SE-0}). Utilizing $\varphi_\alpha^\pm$ defined in \eref{SalpeterWaveFunctionPN},
the equation further reduces two coupled equations:
\begin{gather}
(M-\w_1-\w_2)\varphi_\alpha^+=\frac{\Lambda^+ d_{\alpha\beta} \Gamma^\beta}{2\w_1},\\
(M+\w_1+\w_2)\varphi_\alpha^-=\frac{\Lambda^- d_{\alpha\beta} \Gamma^\beta }{2\w_1},
\end{gather}
which can otherwise be simplified to \eref{E-VS-SE}.

The BS vertex can also be expressed by the Salpeter wave function as
\begin{gather*}
\Gamma^\alpha(P,q)=S^{-1}(p_2)D^{-1}(p_1)\vartheta^{\alpha\beta} \psi_\beta(P,q).
\end{gather*}

\end{document}